# Investigation of microwave loss induced by oxide regrowth in high-Q Nb resonators

J. Verjauw[1,2*], A. Potočnik[2], M. Mongillo[2], R. Acharya[2,3], F. Mohiyaddin[2], G. Simion[2], A. Pacco[2], Ts. Ivanov[2], D. Wan[2], A. Vanleenhove[2], L. Souriau[2], J. Jussot[2], A. Thiam[2], J. Swerts[2], X. Piao[2], S. Couet[2], M. Heyns[1], B. Govoreanu[2] and I. Radu[2]

[1]*Department of Materials Engineering (MTM), KU Leuven, Leuven, B-3000, Belgium*
[2]*imec, Kapeldreef 75, Leuven, B-3001, Belgium,* * jeroen.verjauw@kuleuven.be
[3]*Department of Electrical Engineering (ESAT), KU Leuven, Leuven, B-3000, Belgium*
** corresponding author*

The coherence of state-of-the-art superconducting qubit devices is predominantly limited by two-level-system defects, found primarily at amorphous interface layers. Reducing microwave loss from these interfaces by proper surface treatments is key to push the device performance forward. Here, we study niobium resonators after removing the native oxides with a hydrofluoric acid etch. We investigate the reappearance of microwave losses introduced by surface oxides that grow after exposure to the ambient environment. We find that losses in quantum devices are reduced by an order of magnitude, with internal Q-factors reaching up to $7\times10^6$ in the single photon regime, when devices are exposed to ambient conditions for 16 min. Furthermore, we observe that $Nb_2O_5$ is the only surface oxide that grows significantly within the first 200 hours, following the extended Cabrera-Mott growth model. In this time, microwave losses scale linearly with the $Nb_2O_5$ thickness, with an extracted loss tangent $\tan\delta_{Nb2O5} = 9.9\times10^{-3}$. Our findings are of particular interest for devices spanning from superconducting qubits, quantum-limited amplifiers, microwave kinetic inductance detectors to single photon detectors.

## I. INTRODUCTION

Microwave loss originating from material defects is a major contributor to decoherence and energy relaxation of superconducting quantum devices [1], [2]. Today development of large-scale quantum computers is limited by surface losses where increased device complexity introduces additional lossy interfaces in compact chip designs [3]–[6]. The main contribution to microwave loss at deep cryogenic temperatures (<100mK) and single photon level microwave powers (<fW) is attributed to two-level-system (TLS) defects located at amorphous substrate-metal, metal-air and substrate-air interfaces [1], [2], [7]. By device design, fabrication process optimization and by using high quality materials, microwave loss in superconducting devices has been significantly reduced over the last decade [1], [8], [9]. Nevertheless, the major components of the loss in state-of-the-art devices could be attributed to native oxides that grow on various materials immediately after samples are exposed to ambient air [8]–[10]. Material selection for superconducting devices therefore plays an important role to minimize the microwave loss.

Nb is widely used to fabricate superconducting devices [8], [11]–[15]. Its advantage over other superconducting metals (e.g. Al, TiN, NbN or NbTiN) are low surface roughness [16] and low kinetic inductance [17] associated with reduced device variability and a high superconducting energy gap associated with a lower quasi-particle creation probability [18]. Nb is compatible with industrial level processing [19], its chemical stability yields lower wet etch rates and better etch selectivity towards oxides [12]. Nb is also used to fabricate wafer-scale trilayer Josephson junctions [13]–[15], making it an attractive candidate for large-scale superconducting device integration.

Despite favorable physical and electrical properties, Nb forms 5-7 nm of native oxide, consisting of three components: NbO, $NbO_2$, $Nb_2O_5$ [20], [21]. While the loss tangent of Nb oxide has been studied [22], [23], the contribution of these components to the microwave loss and their evolution during oxide growth is poorly understood. Furthermore, limited strategies for mitigating the losses have been explored [24].

In this work, we study the microwave loss contribution of Nb oxides in superconducting Nb lumped-element resonators (LER) devices. The LER receive a hydrofluoric acid (HF) etch to remove surface oxides [10], followed by a controlled oxide regrowth. Our results show that $Nb_2O_5$ is the dominant source of microwave loss in Nb-based superconducting devices and that it is the only oxide that grows for the first ~200 h. Within this time, we find that microwave loss scales linearly with $Nb_2O_5$ oxide thickness, allowing us to estimate its loss tangent tanδ = $9.9\times10^{-3}$. We conclude that losses can be greatly reduced with the HF etch before the measurement due to the relatively slow logarithmic $Nb_2O_5$ regrowth. Our demonstrated method could be extended to study and remove oxide losses in other superconducting devices.

## II. DEVICE DESCRIPTION

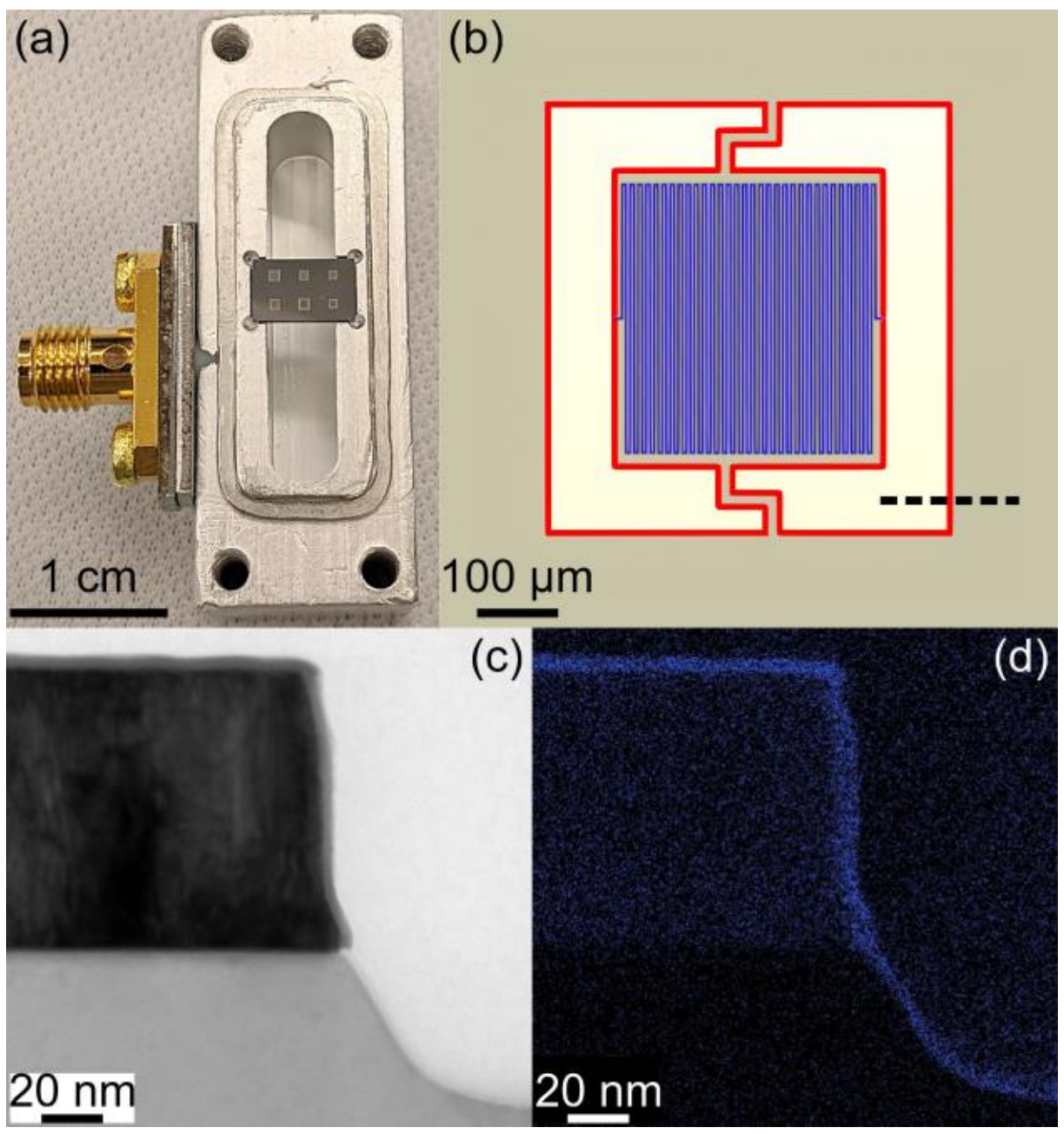

Fig. 1 Sample description. (a) Photograph of a chip containing six LER inside a 3D cavity enclosure. (b) False-colored micrograph of a single LER device comprising of a meander inductor (blue) and capacitor (red). Black dashed line indicates the line-cut of cross-TEM (c) and EDS (d) measurements. (c) Cross-TEM of the LER without HF post-treatment. Oxides are visible at the metal-air and substrate-air interface. (d) EDS visualization of the oxygen content, confirming the presence of the oxides at the above-mentioned interfaces. No oxide is detected at the substrate-metal interface.

Nb lumped-element resonator devices (Fig. 1) are fabricated in the state-of-the-art 300 mm fabrication facility at imec, Belgium. High resistivity silicon wafers (>3 kΩ·cm) first receive an HF clean to remove any native oxides from the surface. To minimize $SiO_x$ regrowth, 100 nm of Nb is deposited at room temperature shortly after the HF clean by physical vapor deposition. The device structures are patterned afterwards using optical lithography and a chlorine-based reactive ion etch. On the patterned wafers, native Nb oxide at the metal-air interface and $SiO_x$ at the substrate-air interface form due to air exposure. These oxides are detected with high-resolution cross transmission electron microscopy (TEM) and energy dispersion spectroscopy (EDS), shown in Fig. 1 (c)-(d), respectively. The measured thickness of native $NbO_x$, $SiO_x$ and the amorphous substrate-metal interface layer are 5.0 ± 0.3 nm, 2.2 ± 0.1 nm and 2.1 ± 0.1 nm, respectively (Appendix A), in agreement with literature [25]–[27]. EDS measurements indicate that the metal-substrate interface layer does not contain oxygen, implying that there is no oxide growth on the substrate before metal deposition. The Nb thin films used for the resonators have been characterized separately using Hall bar structures [34], yielding a critical temperature $T_c$ = 9.02 K, a critical magnetic field $H_c$ = 1.08 T and residual-resistivity ratio $RRR = R(300K)/R(10K) = 6.48$. The Nb film has a compressive stress of 500 MPa and surface roughness of $R_{av}$ = 0.46 nm.

The effects of surface oxides on dielectric loss are studied by initially submerging identically fabricated devices in a 10 vol.% HF solution for 60 s, followed by a deionized water rinse. This procedure removes all present surface oxides. The samples are then subject to a time-controlled oxidation by exposure to cleanroom ambient conditions, which allows us to investigate the intrinsic microwave loss in the resonator as a function of reoxidation time. A sample without HF treatment is also measured and is used as a reference. Additional fabrication, characterization and postprocessing information is found in Appendix A.

## III. OXIDE REGROWTH

Native oxide regrowth is studied on Nb coated wafers and bare silicon wafers with angle resolved x-ray photoelectron spectroscopy (AR-XPS). To mimic the Si surface after LER device patterning, a bare silicon wafer is coated with a Nb layer, which is subsequently etched away. Native Nb and Si oxides are removed and regrown on these samples in the same way as for the LER samples. After the reoxidation period, samples are transferred into a high vacuum environment where XPS is performed. XPS spectra show that Nb, NbO, $NbO_2$ and $Nb_2O_5$ are present at the surface (Fig. 2), in agreement with literature [21]. Spectroscopic peaks corresponding to $Nb_2O_5$ are noticeably higher in the sample with longer reoxidation time, indicating that $Nb_2O_5$ has the highest growth rate. The Nb oxide thicknesses as a function of reoxidation time are calculated from AR-XPS spectra using the $Nb_2O_5/NbO_2/NbO/Nb$ model stack [28] and are shown together with the $SiO_x$ thickness in Fig. 3. Within the first 200 h after oxide removal, $Nb_2O_5$ grows substantially, while the other oxides are still less than one monolayer thick [29], [30]. In this time window, the growth of $SiO_x$ is hindered by the hydrogen-passivated surface from the HF clean [31], [32]. We measure a weak linear increase in $SiO_x$ content, which is less than one monolayer thick for the first $10^2$-$10^3$ h (Fig. 3) [33].

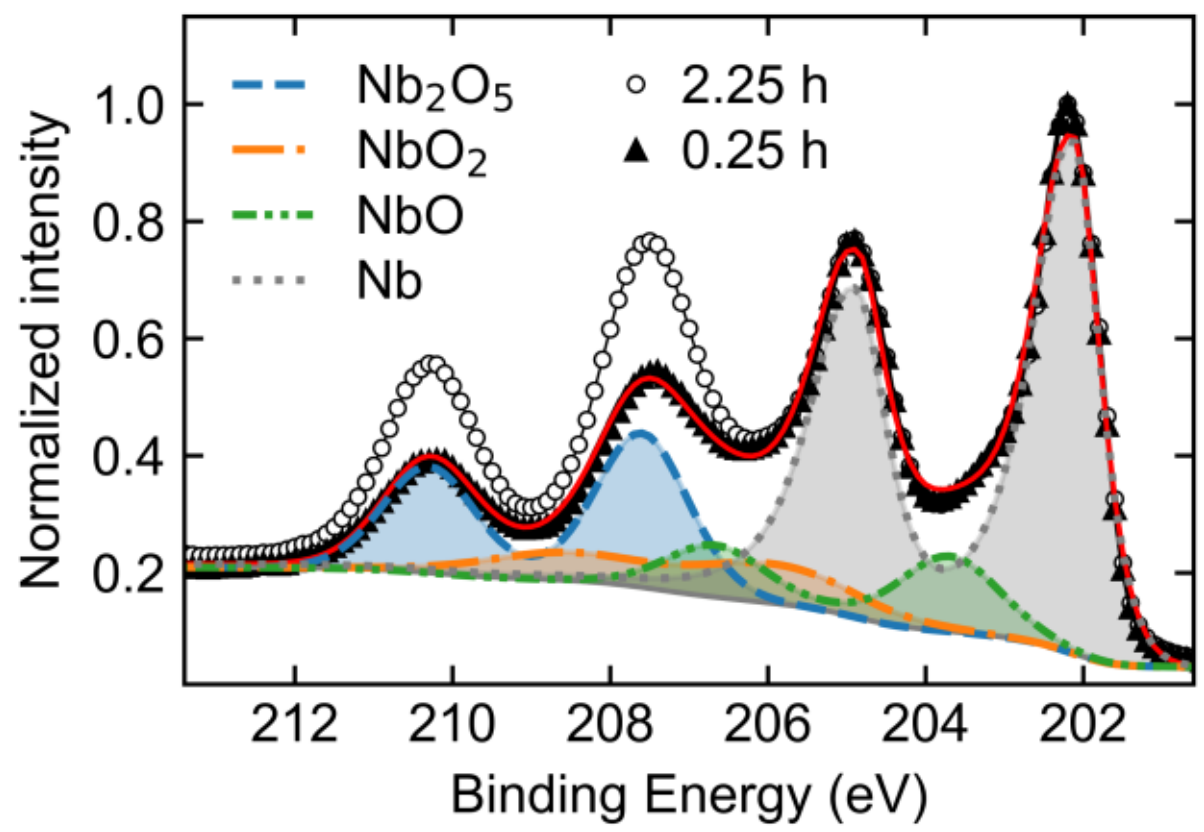

Fig. 2: Angle integrated XPS spectra for Nb coated wafers with short (0.25 h) and long (2.25 h) reoxidation times. The spectra are normalized to simplify chemical comparison. The position of the Nb3d doublets related to oxidized Nb are shown for the data with the shortest reoxidation time.

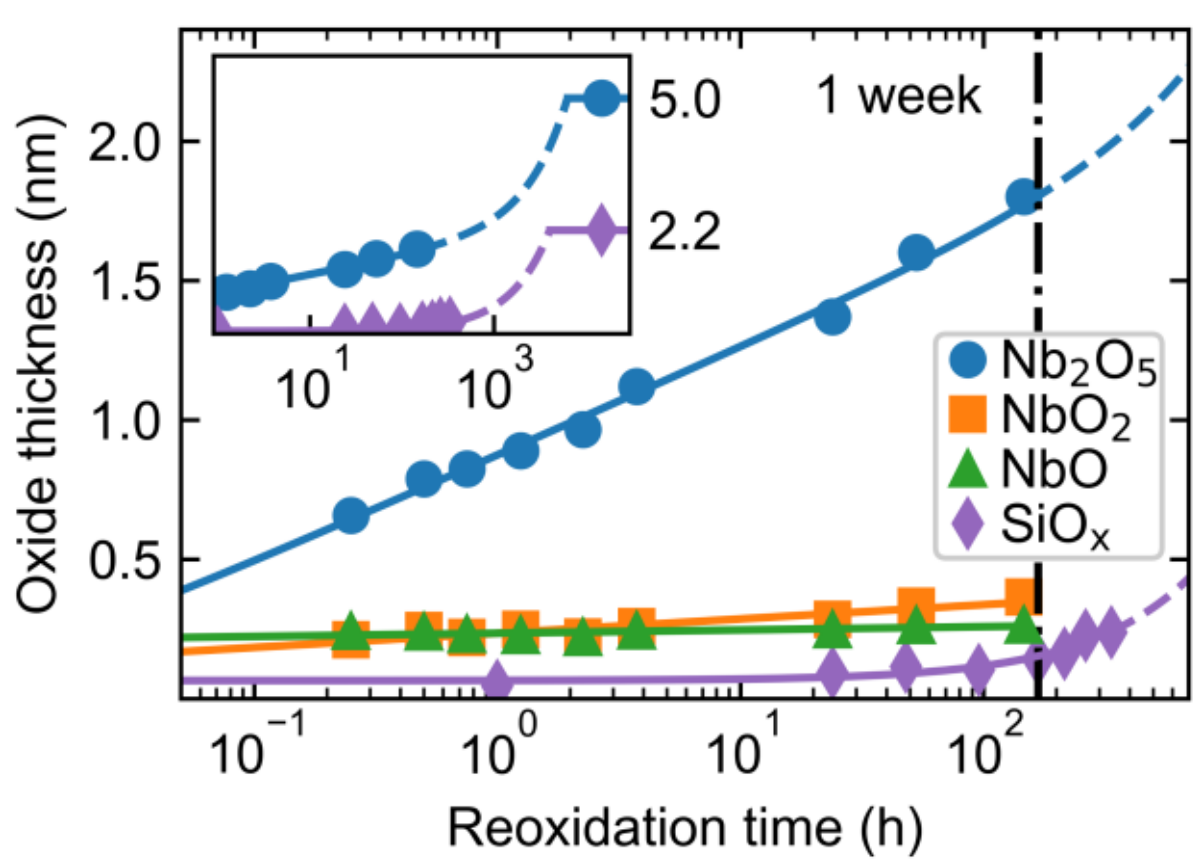

Fig. 3: Si oxide and Nb oxide thicknesses after the HF etch as a function of reoxidation time. Within the first week, only $Nb_2O_5$ grows considerably. Solid lines are fits of the logarithmic CM model for Nb oxide data and linear model for SiOx data. Insert shows thicknesses on a longer timescale and include reference values from TEM measurements. Dashed lines visually connect fits to reference values.

The $Nb_2O_5$ growth follows the extended Cabrera-Mott (CM) model in the electron tunneling limited regime [34] and is described by the logarithmic dependence $d = C \cdot \ln(\alpha t + 1)$, where $d$ is the oxide thickness, $t$ is time and $C$ and $\alpha$ are fitting parameters. For longer reoxidation times, we arbitrarily extend the fit to match thicknesses of the reference sample. This regime, indicated with dashed lines in Fig. 3, is not used for later analysis. Based on literature values we infer that all oxide thicknesses saturate after $10^3$-$10^4$ h [26], [35]. In our experiment, this time corresponds to the age of the reference sample at the time of measurement. We would like to stress that within the measurement uncertainty, $Nb_2O_5$ thickness is the only varying time-dependent parameter for a duration of up to ~200 h.

## IV. MICROWAVE LOSS MEASUREMENTS

To determine the effect of oxide removal and regrowth on superconducting devices, we characterize high-quality LER samples with various reoxidation times. A sample with 6 LER's is placed inside a high-purity aluminum 3D cavity (Fig. 1(a)). Wireless coupling between the resonator and 3D cavity allows for fast sample preparation without the need for gluing and wire bonding. With this method the sample cooldown can start within 16 min (0.27 h) after oxide removal. The LER are characterized at 10 mK and low (<fW) microwave powers, which is the typical operating regime for superconducting quantum devices.

LER resonances are well fitted with a generalized Lorentzian function (Appendix C), from which the loaded quality factor $Q_L$ is extracted. $Q_L$ depends on the internal ($Q_i$) and external ($Q_e$) quality factor: $1/Q_L = 1/Q_i + 1/Q_e$, however we note that in our experiment the measured loaded quality factor of the LER is nearly identical to the internal quality factor ($Q_L \approx Q_i$) due to a large designed external Purcell Q-factor $Q_e \gg Q_i$ (Appendix C). The microwave loss in the device ($\propto 1/Q_i$) can be therefore determined from $Q_L$ directly.

The measured $Q_L$ decreases with decreasing incident power (Fig. 4), which is characteristic of TLS microwave loss [1]. We express the incident microwave power in terms of average photon occupation $\langle n \rangle$ in the LER. The Q-factor power dependences can be well fitted by the two-component TLS loss model [36]–[38]:

$$\frac{1}{Q_\mathrm{L}} \approx \frac{1}{Q_\mathrm{i}} = \sum_{i=1}^{2} \frac{F_i \tan\delta_i}{\left(1 + \frac{\langle n \rangle}{n_{\mathrm{c},i}}\right)^{\beta_i}} + \frac{1}{Q_\mathrm{r}}, \qquad (1)$$

where $F_i$ is the participation ratio and $\tan\delta_i$ the intrinsic loss tangent for the respective $i$-th sub-volume component containing TLS. Similarly, $n_{c,i}$ is the critical photon number equivalent to the saturation field of the different TLS and $\beta_i$ is a phenomenological parameter which is 0.5 for non-interacting TLS defects [39] and lower than 0.5 in the presence of TLS-TLS interactions [40]. $1/Q_r$ is a residual power-independent loss term.

Individual parameters of all sub-volumes $i$ in Eq. 1 cannot be determined accurately due to significant time dependent variations observed in measured Q-factors (Appendix C), as is common in superconducting devices [41]–[45]. However, different trends in the extracted parameters can be observed (Appendix C). The two components most notably differ in critical photon number, with $n_{c,1}$ = 1-$10^2$ and $n_{c,2}$ = $10^5$-$10^7$ for all resonators, giving rise to the observed plateau at $n$ = $10^2$-$10^4$ in Fig. 4. $\beta$'s are generally less than 0.5, indicating that TLS-TLS interactions are present for both components, in agreement with [23]. We note that time dependent Q-factor variation is larger than on-chip resonator-to-resonator and chip-to-chip variation (Appendix C). Within this uncertainty, we also do not observe any prominent Q-factor frequency dependence, nor any dependence of LER resonance frequencies on the reoxidation time (Appendix C). To capture the TLS contribution, we limit further analysis to $Q_L$-factors in the single photon regime, which are denoted as $Q_{L,spr}$.

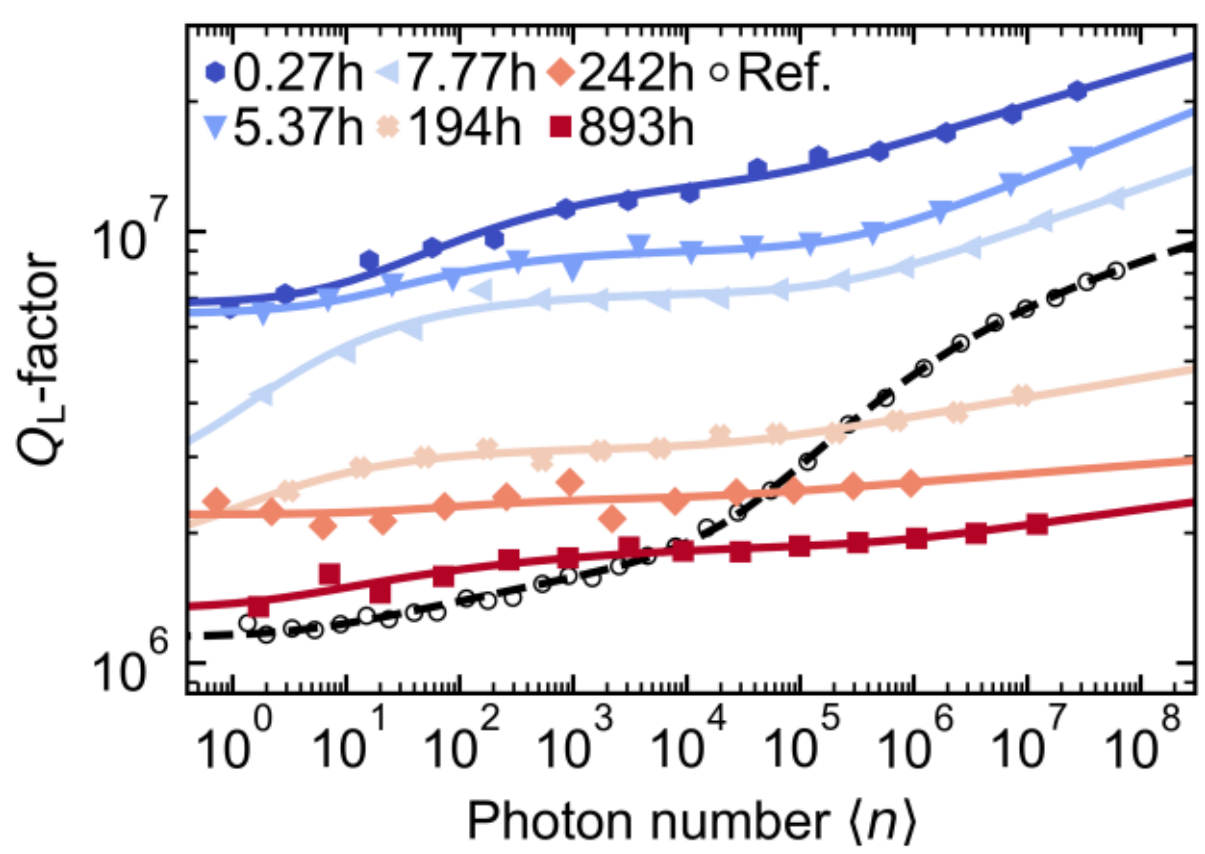

Fig. 4: Power dependent Q-factors of selected LER3 (Appendix C) shown for different reoxidation times. The spectra are fitted by Eq. 1. Reference sample measurements are denoted by empty circles. After oxide removal, Q-factors initially increase relative to the reference sample and decrease with increasing reoxidation.

The reference sample with intact native oxides exhibits $Q_{L,spr} \approx 1\times10^6$. Immediately after oxide removal, the quality factors increase by almost an order of magnitude, reaching $Q_{L,spr} \approx 7\times10^6$ in the sample with the shortest reoxidation time (0.27 h). For longer reoxidation times, low power quality factors gradually return to the reference value. This reaffirms the negative effect of native oxides and the importance of oxide removal [10], [38]. High power Q-factors however fall below the reference, indicating an increase in losses in this regime.

To determine the origin of the different microwave loss contributions, we plot the inverse of all $Q_{L,spr}$-factors as a function of reoxidation time and compare these with the estimated loss from the $Nb_2O_5$, $SiO_x$, and reoxidation time independent component (TIC) associated either with the silicon substrate or the substrate-metal interface (Fig. 5). The loss contributions from NbO and $NbO_2$ are neglected due to their small thickness as well as their superconducting [46] and metallic behavior [47], respectively. The loss components are calculated using Eq. 1 at low photon numbers ($\langle n \rangle / n_{c,i} \approx 0$). The participation ratios, obtained from numerical electromagnetic simulations (Appendix D), scale with the oxide thickness as shown in Fig. 3. The loss tangent of native silicon oxide $\tan\delta_{SiO2} = 1.7\times10^{-3}$ is taken from literature [48], while loss tangent of the $Nb_2O_5$ and loss of TIC are free parameters used to fit the data. Since exact oxide thicknesses are required for time dependent participation ratio estimations, only reference values and values with reoxidation times shorter than 200 h (~1 week) are used in the analysis.

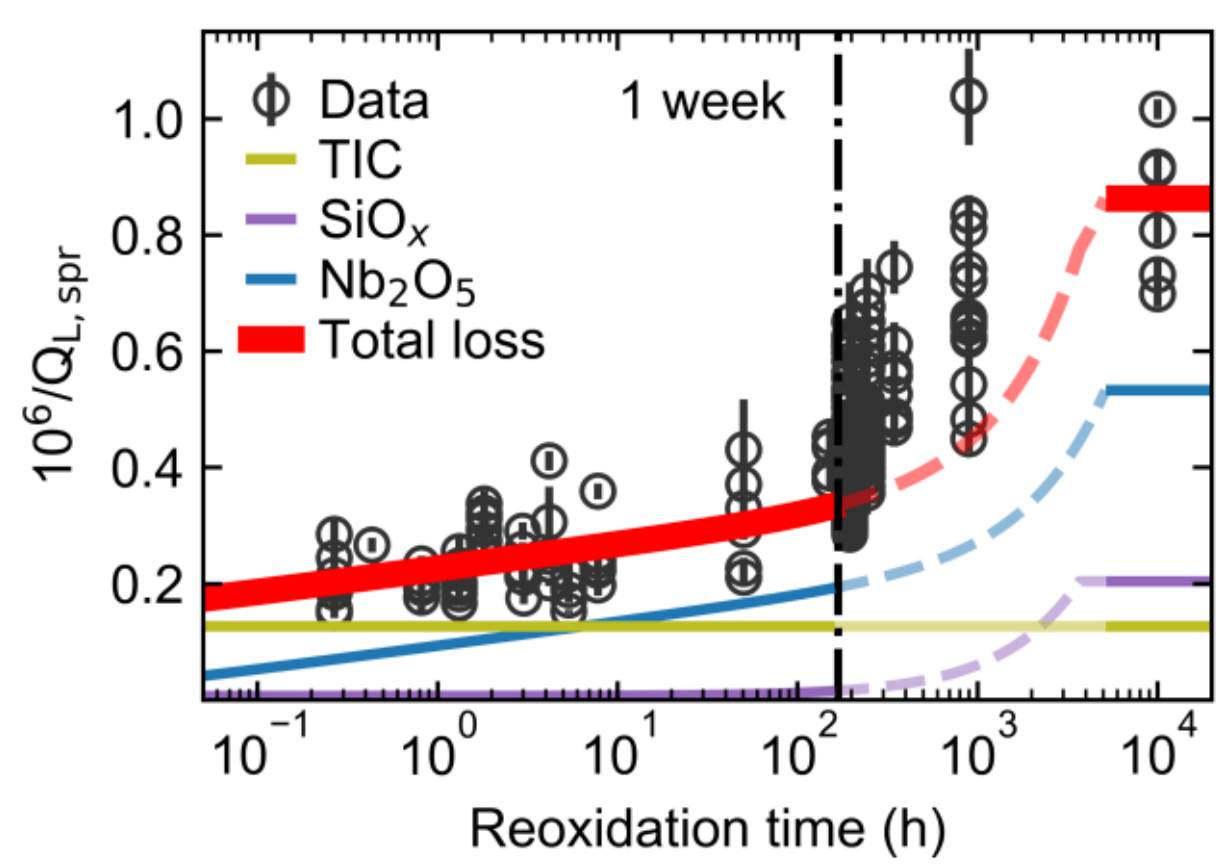

Fig. 5: Single photon limit microwave loss ($1/Q_{L,spr}$) as a function of reoxidation time for all measured LER. The data is fitted using Eq. 1, oxide thicknesses and corresponding calculated participation ratios. Beyond $t$ = 168 h (1 week), the dashed line serves as a guide to the eye for the losses that saturate at the reference sample value.

Good agreement between the calculated loss and the data is obtained for $\tan\delta_{Nb2O5} = (9.9 \pm 0.6)\times10^{-3}$ and TIC loss $F_{TIC}\tan\delta_{TIC} = (1.2 \pm 0.1)\times10^{-7}$. The extracted loss tangent of $Nb_2O_5$ is in agreement with previous work [49]. From the individual calculated loss components (Fig. 5), we observe that for reoxidation times up to ~ 6 h the loss is dominated by the time independent component. Based on simulations, we note that this component can be attributed to either the Si substrate with $\tan\delta_{Si} = (1.3 \pm 0.1)\times10^{-7}$, the metal-substrate interface with $\tan\delta_{M\text{-}S} = (3.5 \pm 0.3)\times10^{-4}$ or possibly a combination of the two. Both loss tangents agree with previously reported values [48]. To identify which of the two gives rise to the time-independent component, we measure a sample with a higher substrate resistivity (>6 kΩ·cm) and with a short reoxidation time (0.35 h). Despite the large uncertainty, two resonators show notably higher Q-factors, up to $Q_{L,spr}$ = 9.4×10$^6$, compared to all other measured LER's (Appendix C). Assuming that the metal-substrate interfaces are identical between the different substrate s, we can infer that the TIC is associated with losses from the Si substrate.

Microwave losses in Nb high-Q resonators become dominated by TLS defects from $Nb_2O_5$ after ~6 h following the oxide removal, while $SiO_x$ starts contributing to the loss only after ~200 h. In the fully oxidized reference sample, $SiO_x$ represents 24% of the total microwave loss, $Nb_2O_5$ 62% and the substrate 14%. From these results we can conclude that the dominant loss mechanism in Nb superconducting devices is attributed to $Nb_2O_5$, which can be successfully removed with an HF etch. Due to logarithmic $Nb_2O_5$ regrowth, losses double ($Q_{L,spr} \approx 4\times10^6$) in ~10 h compared to loss observed after immediate sample cooldown within 0.27 h ($Q_{L,spr} \approx 7\times10^6$).

Finally, we discuss the origin of the two-components observed in the power dependent Q-factor measurements (Fig. 4). Since $Q_L$ notably reduces with increasing reoxidation time at high photon numbers, and since the $Nb_2O_5$ thickness is the main parameter that changes as a function of reoxidation time, we can conclude that the loss component with the largest critical photon number $n_{c,2}$ is associated with $Nb_2O_5$. The other component with $n_{c,1} \approx 1 - 10^2$, must therefore originate from the TIC. Furthermore, the critical photon number $n_{c,2} \approx 10^5$ for the reference sample increases up to $n_{c,2} \approx 10^7$ after oxide removal (Appendix C). The change of $n_{c,2}$ after oxide removal indicates that TLS defects in the regrown Nb oxide differ from defects in the original oxide, which is also reflected by the discrepancy in high photon number Q-factors between samples with long reoxidation times and the reference. This is corroborated with an increase in surface roughness of the regrown oxide after HF clean, compared to the roughness before HF exposure (Appendix E).

## V. CONCLUSION

We demonstrated that a post-fabrication surface clean with HF acid can effectively reduce microwave losses by almost an order of magnitude in Nb-based superconducting high-Q resonators. We found that after oxide removal, $Nb_2O_5$ is the only growing oxide within the first 200 h. The $Nb_2O_5$ grows logarithmically and microwave losses in the device scale linearly with this oxide thickness. We extracted the $Nb_2O_5$ loss tangent $\tan\delta_{Nb2O5} = 9.9\times10^{-3}$. Our demonstrated procedure involving a combination of a well-controlled oxide regrowth, different material characterization techniques and measurement of resonator quality factors can be extended to selectively study other metal oxides losses relevant for quantum devices. Furthermore, the demonstrated oxide removal method can be utilized to reduce microwave losses in other devices, such as superconducting qubits, quantum-limited amplifiers [50], microwave kinetic inductance detectors [51], single photon detectors [52] or cryogenic filters [53].

## ACKNOWLEDGEMENTS

The authors gratefully thank the imec P-line, operational support and Paola Favia, Hugo Bender and Chris Drijbooms for metrology support. This work was supported in part by the imec Industrial Affiliation Program on Quantum Computing. The authors would also like to thank prof. A. Wallraff for providing lumped element resonator designs.

## APPENDIX A: FABRICATION AND CHARACTERIZATION

Resonators are fabricated in an integrated 300mm pilot line: high-resistivity silicon wafers first receive an HF treatment to remove native silicon oxides from the surface. Immediately after, a 100 nm thick niobium layer is deposited with very high uniformity using DC magnetron sputtering. The device structures are patterned in the Nb layer by using a $SiO_x$ hardmask and a chlorine based reactive ion etch. After patterning, the device wafer is cleaned with an organic solvent and HF solution. This removes any remaining surface and hardmask residues, while the wet etch rate of Nb was found to be negligible. The profile and precise taper angle are controlled by the etch chemistry and use of hardmask. Note that the samples are over-etched, which creates trenching in the silicon substrate to ensure full clearance of the metal film and reduces losses from substrate the metal interface [54]. Native Nb and Si oxides are removed by submerging a diced sample in a 10 vol.% HF for 60 s, followed by a 60 s DIW rinse and $N_2$ blow dry. Over 20 identical samples are post-processed with this procedure and stored in the cleanroom (T: 20°C ± 1 °C, RH: 40% ± 2.5%) for a variable amount of time.

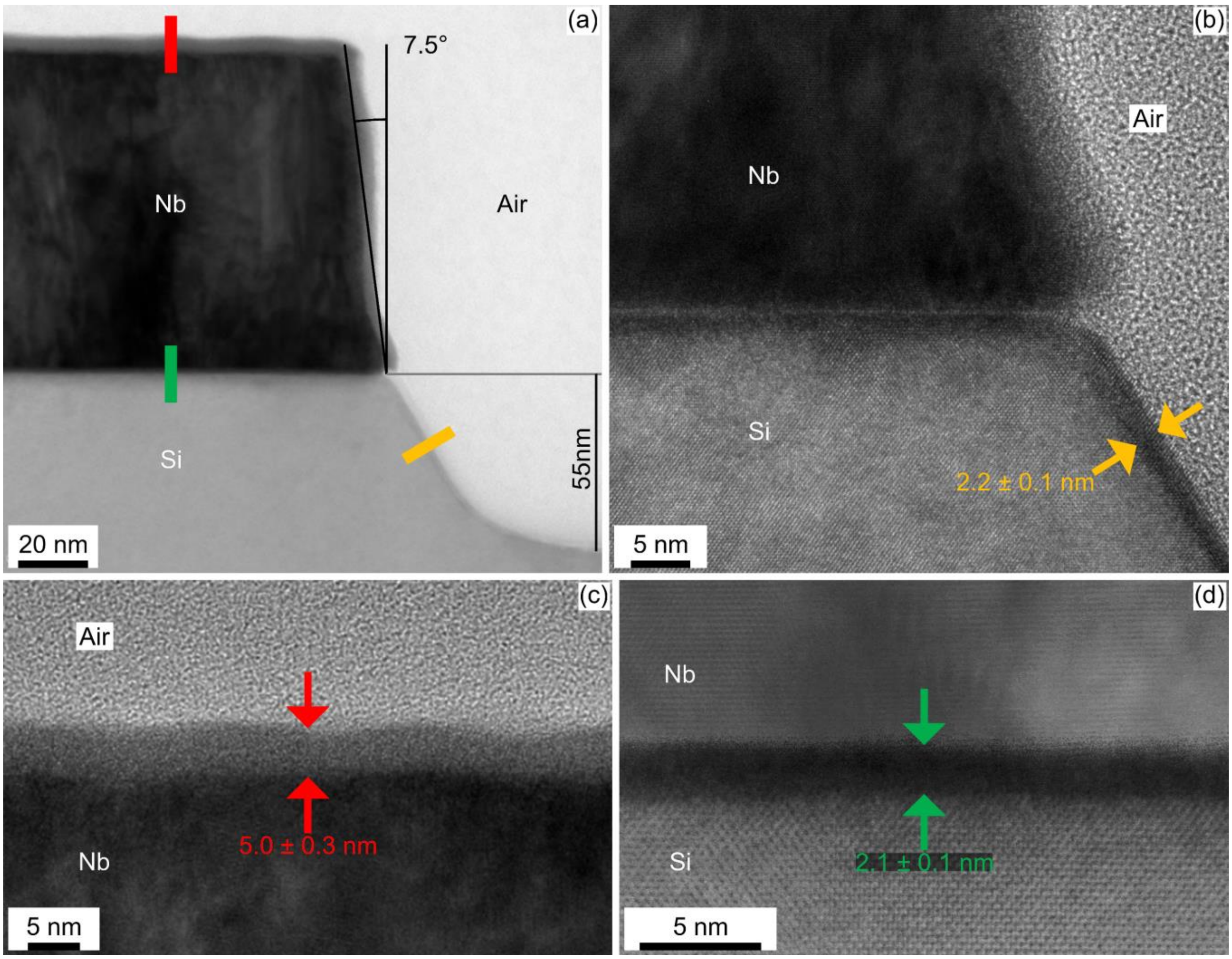

Fig. S1: High resolution cross-TEM image of the different interface layers. (a) Cross-TEM at LER edge without HF treatment (dashed line in Fig. 1 (b)). A sidewall slope of 7.5° and silicon recess of 55nm are observed. (b) Metal-air interface: $NbO_x$ is seen at the interface with a thickness of 5.0 ± 0.3 nm. (c): Substrate-air interface: $SiO_x$ is seen at the interface with a thickness of 2.2 ± 0.1 nm. (d): Metal-substrate interface: while no oxide has formed, an amorphous interface layer of 2.1 ± 0.1 nm is present.

The characterization of the niobium used in this experiment was performed previously [55] and is summarized in Table S-I. Additionally, the cross-section of resonator reference samples is analyzed with cross-transmission electron microscopy (cross-TEM) (Fig. S1). Cross-sections reveal an overall sidewall slope of 7.5° and silicon recess of 55 nm. From high resolution TEM pictures, native oxide thicknesses at the metal-air and substrate-air interface are determined, which are reported in the main text.

TABLE S-I: Material characterization

| $T_c$ (K) | $B_c$ (T) | RRR | $\rho$ (μΩ · cm) | roughness (nm) | | |
|---|---|---|---|---|---|---|
| | | | | range | $R_{av}$ | RMS |
| 9.02 | 1.08 | 6.48 | 19.35 | 4.7 | 0.46 | 0.58 |

To study native oxides with XPS, two sets of samples are prepared. For Nb oxides, unpatterned Nb device wafers are used. To study Si oxides we mimic the Si surface after device patterning by fully removing Nb from a Nb coated wafer using the same etching step as is used for LER devices. To remove the native oxides, the samples were submerged in a 10 vol.% HF solution for 10 minutes, followed by a 1 min DIW rinse and $N_2$ dry. Afterwards, the samples are left in cleanroom ambient for varying aging times, during which the oxides regrew. Finally, all samples are transported to the XPS measurement setup in a vacuum box and then transferred to the high vacuum XPS chamber.

## APPENDIX B: AR-XPS measurement and thickness extraction

**Niobium oxides:**

To extract the Nb oxide thicknesses, angle resolved X-ray photoelectron spectroscopy measurements are carried out on a Theta 300 tool of Thermo Fisher. This system is equipped with a monochromatized Al Ka X-ray source (1486.6 eV). A spot size of 400µm is used. Charge neutralization is used during the measurement. The tool allows parallel angle resolved measurements at 16 angles between 21° and 78° emission angle (angle from the normal of the sample). Relative sensitivity factors are used to convert peak areas to atomic concentrations. As a result of this, it is possible that the concentrations could deviate from reality. Comparison between the atomic concentrations of several samples, however, is more accurate.

Nb metal gives asymmetric Nb3d peak shapes, whereas Nb oxides have symmetric peak shapes. The Nb3d doublet of metallic Nb is positioned at the lowest binding energy position, the one for $Nb_2O_5$ at the highest binding energy position (Table S-II). In between are the doublets (not separately visible on the curves) for NbO and $NbO_2$. The fitting has been executed in the Avantage software of Thermo Fisher Scientific. For the fitting of the asymmetric peaks, a reference Nb3d curve is recorded. After fitting the metallic reference, the tailing parameters are fixed as much as possible and used for the fitting of the oxidized Nb samples, from which the atomic concentrations are extracted. Layer thicknesses are subsequently calculated from these concentrations using the model stack $Nb_2O_5$/$NbO_2$/NbO/Nb.

Table S-II: Nb XPS peaks

| Component | XPS spectral line | Peak binding energy |
|---|---|---|
| Nb | $3d_{5/2}$ | 202.1 eV |
| | $3d_{3/2}$ | 204.8 eV |
| NbO | $3d_{5/2}$ | 203.6 eV |
| | $3d_{3/2}$ | 206.3 eV |
| $NbO_2$ | $3d_{5/2}$ | 205.4 eV |
| | $3d_{3/2}$ | 208.1 eV |
| $Nb_2O_5$ | $3d_{5/2}$ | 207.6 eV |
| | $3d_{3/2}$ | 210.3 eV |

To the extracted Nb oxide thicknesses we fit the Cabrera-Mott model $d = C \cdot \ln(\alpha t + 1)$ (Fig. 3). Extracted parameters are shown in Table S-III. For $Nb_2O_5$, there is good agreement between the data and the model. While the model fits to the NbO and $NbO_2$ oxides, there is a large uncertainty on the fitted values due to the small relative change in the extracted datapoints relative to their uncertainty. When extrapolating the fits to the saturated Nb oxide thickness of 5 nm, we see that $NbO_2$ and NbO have a thickness of 0.63 nm and 0.32 nm, respectively. However due to the large uncertainty, we limit the fit for the available data in Fig. 3 to the last datapoint.

Table S-III: Cabrera-Mott model fit parameters

| Component | C | $\alpha$ |
|---|---|---|
| NbO | 0.005 | $1.188\times10^{20}$ |
| $NbO_2$ | 0.168 | $3.972\times10^{5}$ |
| $Nb_2O_5$ | 0.177 | $1.364\times10^{3}$ |

**Silicon oxide:**
The measurements are carried out in Angle Resolved mode using a PHI Quantes instrument from Physical Electronics. The measurements are performed using a monochromatized photon beam of 1486.6 eV and a spot-size of 100 µm. As for the Nb oxide measurements, instrument specific relative sensitivity factors are used to convert peak areas to atomic concentrations. The measurement has been executed in angle resolved mode, recording XPS spectra are recorded at two take-off angles (TOA, the angle to the sample surface); 45° and 90°. To fit the data only two components have been used: Si (substrate) and $SiO_2$ (Fig. S2 (a)). Due to a very thin $SiO_2$ layer, a lower density of 2.196g/cm$^3$ is used instead of the bulk density 2.65g/cm$^3$ for layer thickness extraction. The calculated layer thicknesses for most samples are extremely low, and therefore the 'Surface concentration' would be a more appropriate parameter.

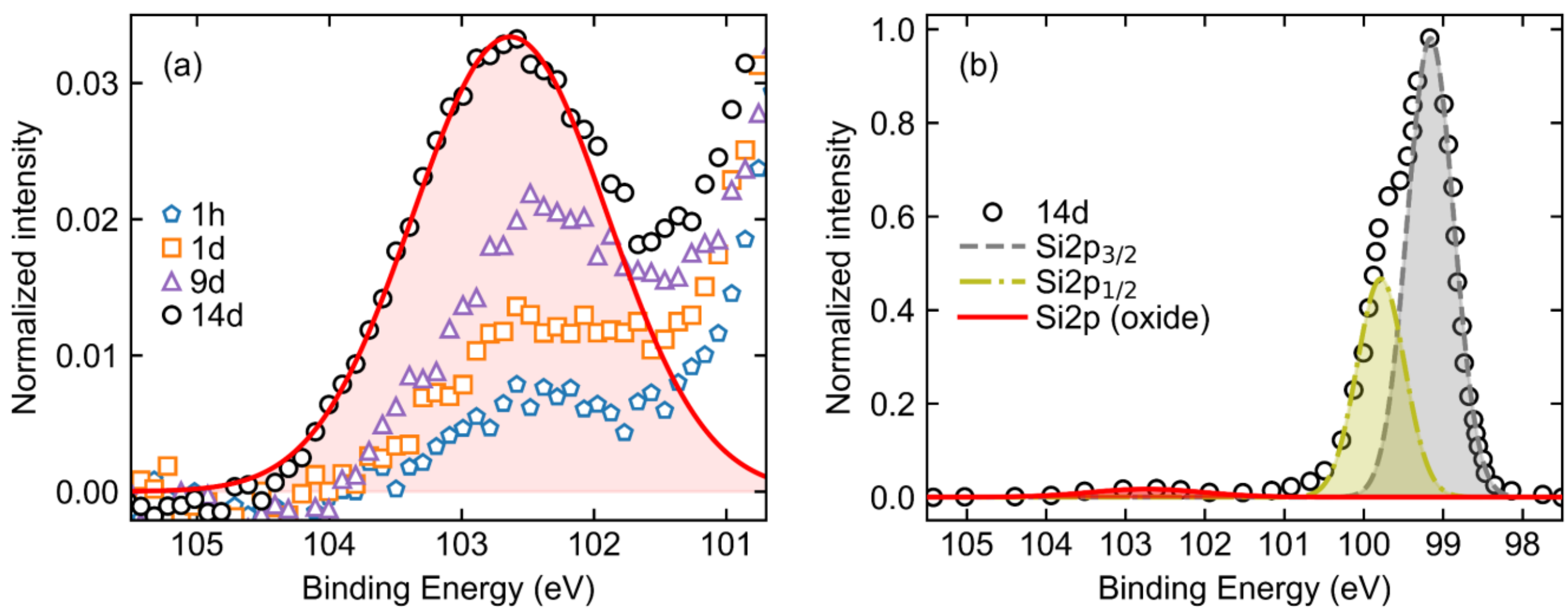

Fig. S2: Si2p XPS spectra of the Si surface (a): Zoom inside the normalized Si2p XPS spectrum on the Si2p peak at a binding energy position typical for $SiO_2$ to compare several reoxidation times. When increasing the reoxidation time, the silicon oxide content fraction increases. The fitting of the Si2p is shown for the longest reoxidized sample (14 days). (b): Fitting of the full Si2p spectrum for the longest reoxidation time: Si substrate Si2p doublet and the $SiO_2$ Si2p peak are located at lower and higher binding energy position, respectively. All spectra in this figure are recorded at TOA 90°.

## APPENDIX C: Q-FACTOR MEASUREMENTS

The performance of superconducting resonators is quantified by their Q-factor (energy stored/energy dissipated per cycle), which is related to the photon lifetime inside the resonator. Q-factors are measured using a sample containing 6 LER's inside a 3D aluminum cavity. The measured transmission spectrum $S_{21}$ shows a TE101 cavity resonance at $\nu_{cav}$ = 7.959 GHz and 6 LER resonances with frequencies $\nu_{LER}$ between 3 and 6 GHz (Fig. S3 (a), Table S-V). For the fundamental mode of each resonator, microwave transmission spectra $S_{21}$ are measured. The magnitude of each spectrum reveals an asymmetric line-shape due to a strong impedance mismatch induced by the coupling antennae of the 3D cavity [56]. The spectra can be well fitted with Eq. C1, from which the loaded quality factors $Q_L$ are extracted (Fig S5 (c)):

$$|S_{21}| = \frac{A(Q_{\mathrm{L}}/|Q_{\mathrm{e}}|)e^{i\phi}}{1 + 2iQ_{\mathrm{L}}(f - f_{\mathrm{r}})/f_{\mathrm{r}}} + B. \qquad \text{(C1)}$$

where $f_r$ is the resonance frequency, and $\phi$ quantifies the impedance mismatch. The environment is accounted for by the parameters $A$ and $B$. The pre-factor $A$ depends on attenuation and gain in the measurement setup. $Q_L$ is determined by the internal ($Q_i$) and external ($Q_e$) quality factor: $1/Q_L = 1/Q_i + 1/Q_e$. The Purcell decay of the resonator to the cavity determines $Q_e$: the LER can decay through the cavity resonance via the Purcell decay. The Purcell decay is quantified by $\kappa_{Purcell} = \kappa_{cav}\,(g/\Delta)^2$ [57]. $\kappa_{cav}$ is the linewidth of the cavity, $\Delta$ is the distance of the LER resonance to the cavity mode and $g$ = 62MHz is the coupling of the LER to the cavity. If the LER is limited by Purcell decay, the measured Q-factor is $Q_L \approx Q_e = \omega_{LER}/\kappa_{Purcell}$. In terms of cavity Q-factor, $Q_e = \omega_{LER}/\omega_{cav} \cdot (\Delta/g)^2 \cdot Q_{cav}$. We see that for the LER closest to the cavity resonance, the Purcell decay ($Q_e$) is largest (smallest). Hence for LER6, $Q_e \approx 1117 \cdot Q_{cav}$. By ensuring that the cavity Q-factor is large ($>10^4$), the measured LER Q-factors are not Purcell limited ($Q_e > 10^7 \gg Q_i$). Therefore, $Q_L \approx Q_i$ and $Q_L$ becomes a direct measure of the resonator's internal quality factor.

Table S-IV: Sample description

| Sample name | Reoxidation time (h) | range ($10^6$) |
|---|---|---|
| A1 | 0.27 | 3.5 - 6.7 |
| B1* | 0.35 | 2.5 - 9.4 |
| A2 | 0.43 | 3.8 - 3.8 |
| A3 | 0.82 | 4.3 - 5.7 |
| A4 | 1.32 | 3.9 - 6.0 |
| A5 | 1.82 | 3.0 - 4.1 |
| A6 | 2.97 | 3.4 - 4.7 |
| A7 | 3.02 | 4.5 - 5.7 |
| A8 | 4.17 | 2.4 - 5.0 |
| A9 | 5.37 | 4.6 - 6.5 |
| A10 | 7.77 | 2.8 - 5.0 |
| A11 | 50.08 | 2.3 - 4.7 |
| A12 | 151.5 | 2.2 - 2.6 |
| A13 | 194.62 | 1.5 - 3.5 |
| A14 | 242.87 | 1.4 - 2.8 |
| A15 | 343.78 | 1.3 - 2.1 |
| A16 | 893.92 | 1.0 - 2.2 |
| A17 | Reference | 1.0 - 1.4 |

* Sample fabricated on 6k wafer. Other samples are fabricated on a single 3k wafer.

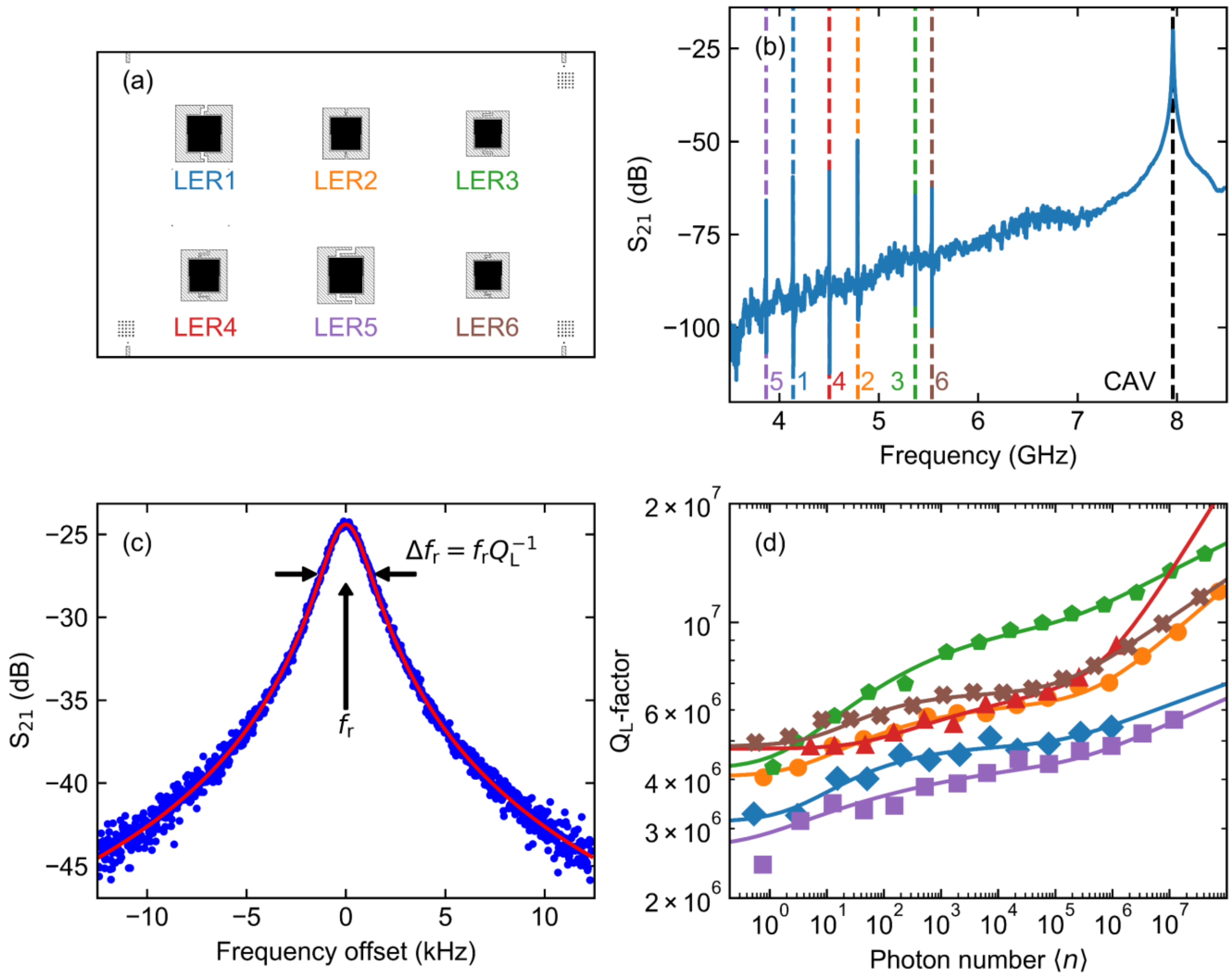

Fig. S3: Device layout and spectroscopy. (a) LER chip design with denoted resonator names. (b) $S_{21}$ scattering parameter measurement over the full frequency range for 6 LER in a 3D cavity. The cavity resonance is found at 7.959 GHz, while the LER resonances lay in the lower extent of the cavity tail. (c) $S_{21}$ measurement near a single LER resonance. By fitting the magnitude with a modified Lorentzian, the resonance frequency $f_r$ and linewidth $\Delta f_r$ are extracted, from which the loaded Q-factor is extracted. (d) Sample A8 Q-factor power dependence (power in terms of photon number).

TABLE S-V: LER resonance frequencies

| Resonator name | LER1 | LER2 | LER3 | LER4 | LER5 | LER6 |
|---|---|---|---|---|---|---|
| Resonance frequency (GHz) | 4.137 | 4.787 | 5.367 | 4.503 | 3.868 | 5.534 |

The resonators are measured for varying input powers, from which power dependent $Q_L$ are extracted (Fig. S3). The data is measured only up to $10^8$ photons in a resonator. Above this value the Duffing effect is observed. Power dependences are fitted using Eq. 1 from the main text. We note that for datasets where Q-factors do not saturate at high power, the power independent term ($Q_0$) is omitted during fitting. This does not affect the other fitting parameters. Furthermore, we limit $\beta$'s between 0 and 0.5. The extracted parameters are shown in Fig. S4, where the components are assigned based on critical photon number. From extracted parameter distributions we observe that two components can be distinguished by critical photon number (Fig. S4 (a)-(b)). We identify the $Nb_2O_5$ contribution with the component which has the largest $n_c$ (see main text). The weighted loss tangents seem randomly distributed between both components with the tendency of $Nb_2O_5$ being lower (Fig. S4 (c)-(d)). Large parameter spread can also be observed for $\beta$. We see that the component corresponding to $Nb_2O_5$ is consistently below 0.5, suggesting that strong TLS-TLS interactions are present in this component. This is in agreement with literature [23]. The second component has higher values, indicating smaller TLS-TLS interaction. We note that distributions of both components are skewed due to limits we set on the fitting model and the absence of clear Q-factor saturation both at high and low photon numbers. For this reason, no quantitative conclusions can be drawn from these results.

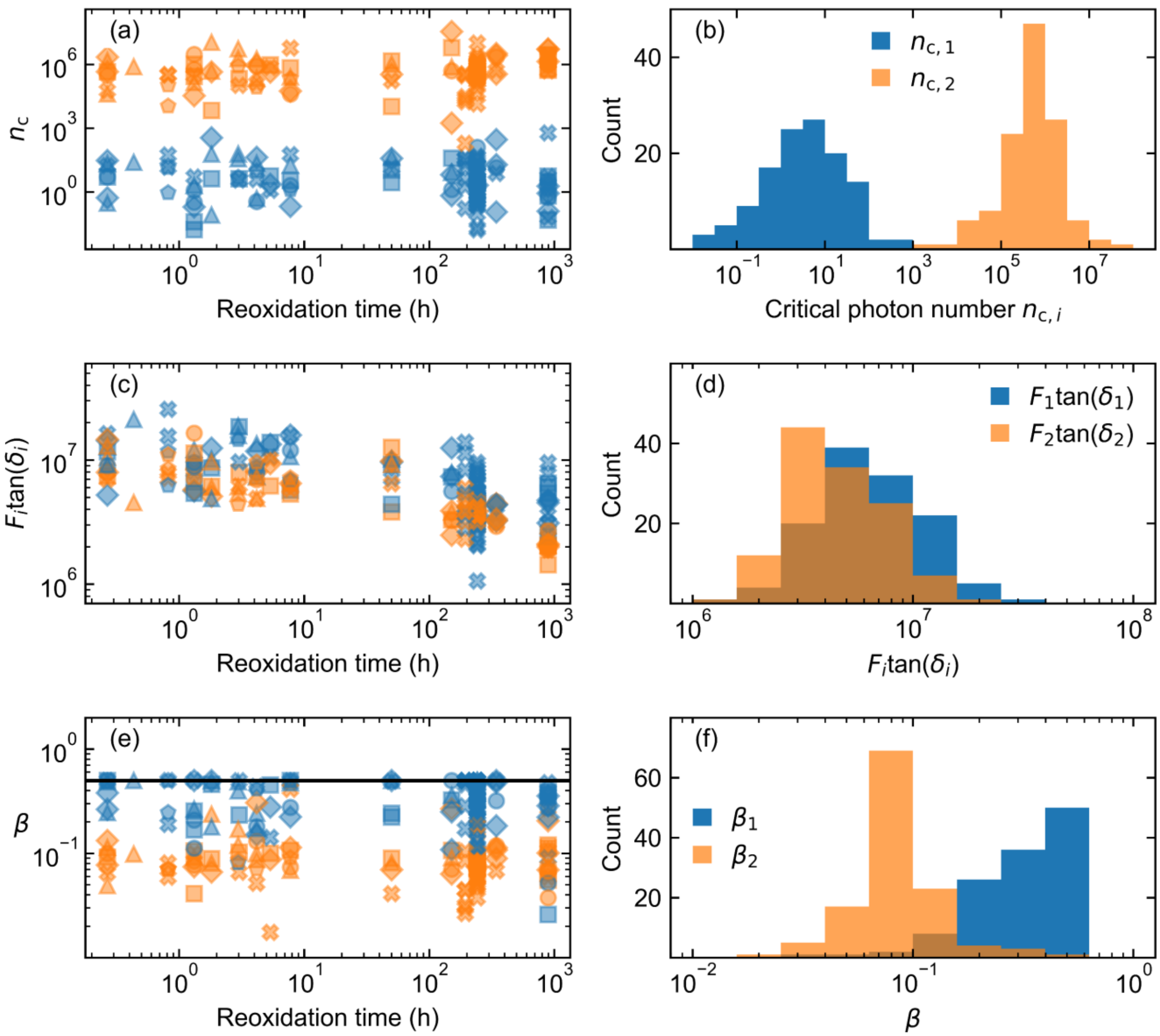

Fig. S4: Extracted fitting parameters from the two-component model (Eq. 1) with logarithmic distributions are shown for (a,b) critical photon numbers, (c,d) weighted loss tangents and (e,f) $\beta$ parameters. Different resonators are denoted as LER1: ●, LER2: ▲, LER3: ■, LER4: ⬟, LER5: ◆, LER6: ✖.

When TLS that are strongly coupled to the LER move in and out of the LER resonance, their Q-factors alter greatly. This is the reason we observe the large spread in fitting parameters above. In order to identify the origin of large parameter spread observed above and to quantify the uncertainty of the low photon number quality factors shown in Fig. 5, we perform repeated Q-factor measurements on samples over an extended period of time. First, we measure photon number dependences of a 5.538 GHz mode for 16 consecutive hours (Fig. S5 (a)). The sample's reoxidation time is 243 h. We observe a switching behavior both at high and low photon numbers. Two separate bands are visible in the high photon number regime, which is in agreement with telegraph noise from TLS, observed in superconducting devices [41]. Secondly, a 5.361 GHz resonator was measured in the single photon regime for 12 days. This sample has a reoxidation time of 195 h. The Q-factor of the LER is normally distributed, with standard deviation of 11% (Fig. S5 (b)).

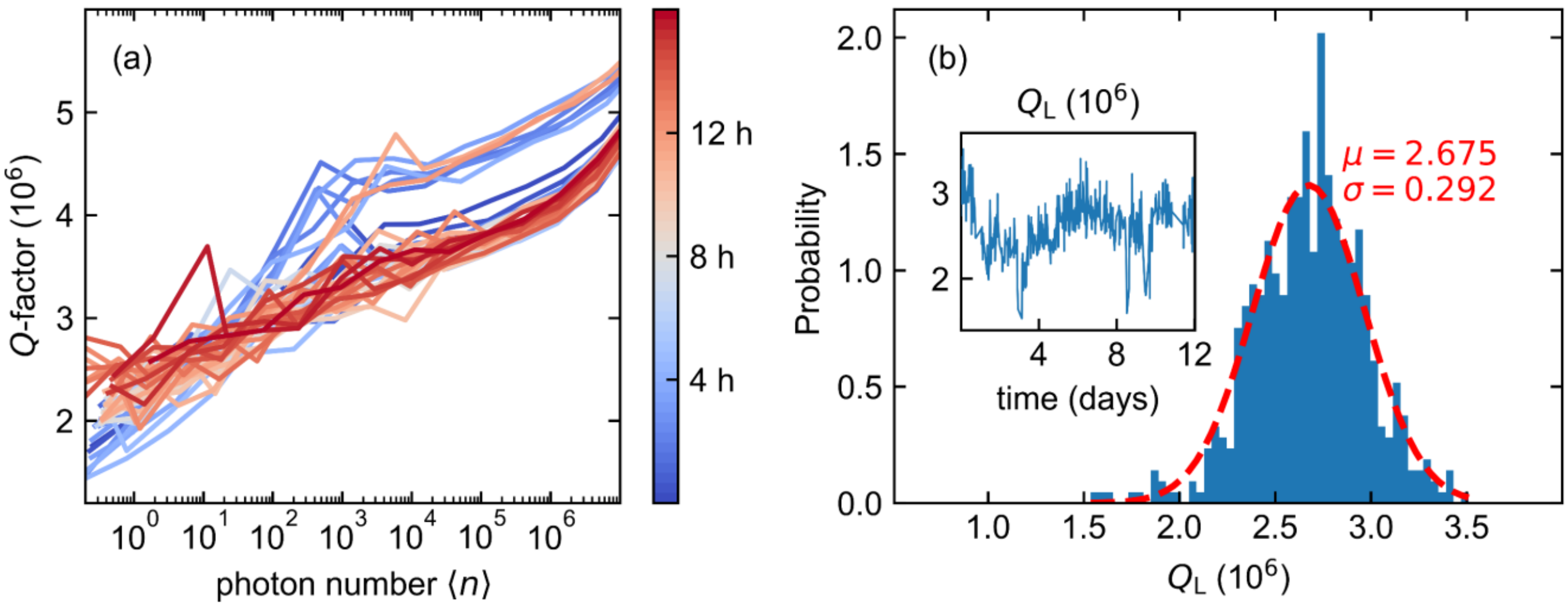

Fig. S5: Time domain measurements. (a): Photon-number dependence of LER resonance (5.538 GHz) was measured for 16 h. Two bands are visible at high photon number, suggesting TLS switching. (b) LER resonance (5.361 GHz) was measured for 12 days, during which $Q_L$ was extracted. The Q-factor has a normal distribution with an uncertainty of 11%. Insert: $Q_L$ fluctuations versus time at single photon level.

Finally, within the measurements uncertainty we do not observe any prominent quality factor dependence on the resonator frequency (Fig. S6 (a)), nor any dependence of the resonator frequencies on the reoxidation time (Fig. S6 (b)). The frequency offset is calculated with respect to the untreated reference sample. A sample with higher substrate resistance (6 kΩ·cm) was also measured. For this sample, we see a strong quality factor dependence, indicating the importance of the substrate contribution to the losses (we consider the metal substrate interface to be the same for both substrates). Furthermore, a stronger frequency shift is observed for this sample.

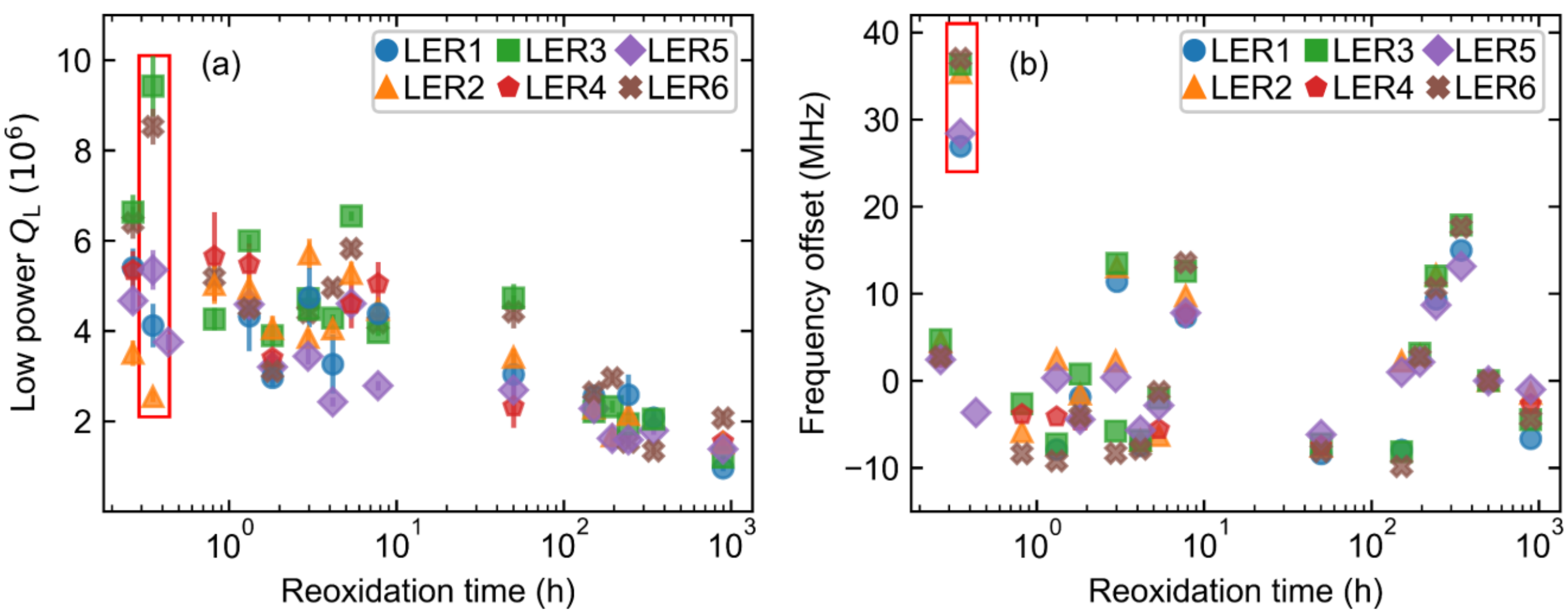

Fig. S6: The (a) quality factor and (b) frequency offset for different reoxidation times of all LER. Low $Q_L$-values are the inverse of datapoints in Fig. 5. We observe a substrate dependence on both quantities (red rectangle).

**APPENDIX D: SIMULATIONS**

The participation ratios ($F$) of the dielectric regions are estimated from electric field distributions, which are obtained from finite-element numerical simulations of the design. Due to the absence of translational symmetry and the large variation (~ nm to several 100 μm) in the length scales associated with the design, a 2-step approach is adopted to obtain the participation ratios. First, a coarse 3D simulation of the LER is performed using the eigenmode solution setup of the high-frequency electromagnetic solver Ansys HFSS [58]. The LER is modeled as a 2D metal sheet on a silicon substrate and is placed in a 3D air box. This enables an accurate simulation of the electric field distribution in the resonator at its resonance frequency. Several sub-regions are defined in the design as shown in Fig. S7 (a) and the electric field distribution is used to estimate the fraction of the total energy present in each sub-region.

Next, a fine 2D simulation of each of the sub-regions is performed by solving the Poisson equation in Sentaurus TCAD [59]. For these simulations, apart from the metal conductors, the appropriate dielectrics are included with thicknesses obtained from physical characterization methods. Figure S7 (b) shows the result obtained for the electric field distribution (normalized) in the capacitance-meander region (CMR). The electric field is subsequently used to estimate the $F_i$ for each dielectric. The simulation is then repeated for the capacitance region (CR) and the meander region (MR). In addition, the TCAD simulations can also be used to obtain the dependence of the participation ratio of the metal-oxide for different oxide thicknesses. As shown in Fig. S7 (c), this participation ratio varies linearly with the oxide thickness.

As a final step, we estimate the total participation ratio of a dielectric as a sum of the participation ratio in various sub-regions, weighted with their corresponding energy fractions obtained from HFSS. The results obtained are summarized in Table S-VI.

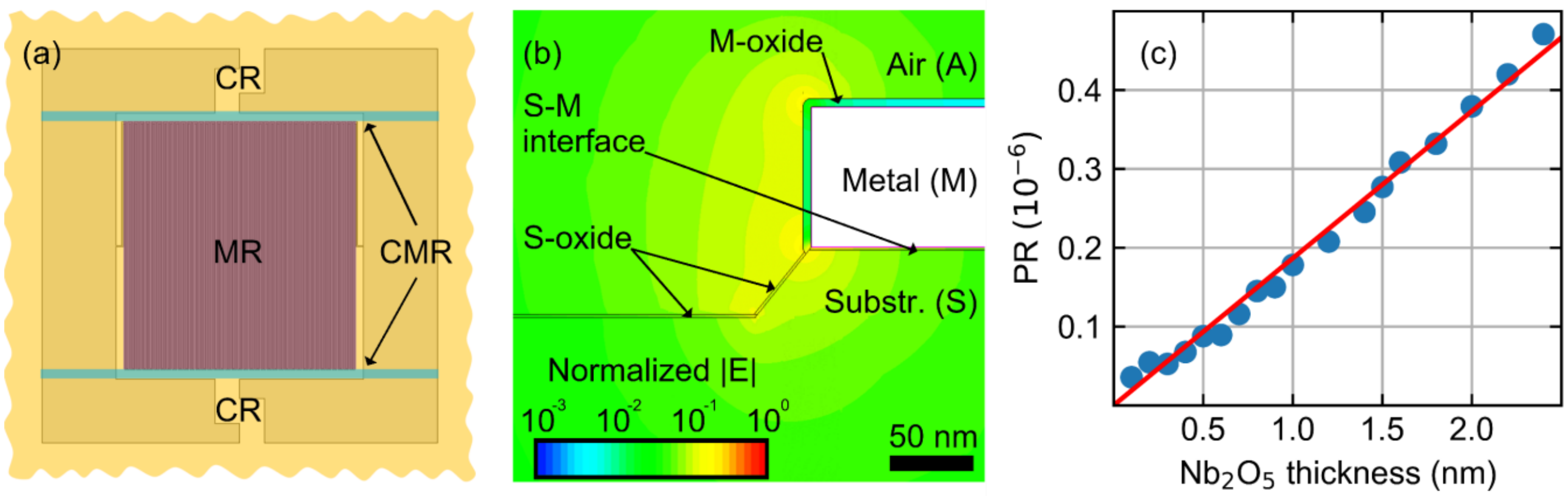

Fig. S7: Simulation results for estimating the participation ratios. (a) Regions identified to estimate the fraction of the total energy using Ansys HFSS. The capacitor region (CR) has ~78.1%, the meander region (MR) has ~16.9% and the capacitance-meander region (CMR) has ~5.1% of the total energy. (b) Normalized electric field distribution obtained from Sentaurus TCAD for CMR around the capacitor conductor. The widths of the meander conductor and the capacitor conductors are taken as 3 μm and 300 μm respectively and have a spacing of 18.5 μm between them. Both conductors have a thickness of 100 nm. The thicknesses of the metal-oxide (M-oxide), the substrate-oxide (S-oxide) and the substrate-metal (S-M) interfacial layer are 5.0 nm, 2.2 nm and 2.1 nm, respectively. Similar simulations have also been performed (not shown here) for the CR (conductor widths 300 μm and spacing 50 μm) and MR (conductor widths 3 μm and spacing 3 μm). For the simulations, the dielectric constants for silicon substrate ($\epsilon = 11.9$), substrate oxide ($\epsilon = 4$), metal-substrate interfacial layer ($\epsilon = 10$) and metal oxide ($\epsilon = 10$) are assumed. (c) Participation ratio of the metal-oxide as a function of the metal-oxide thickness for CR. It is seen that the participation ratio varies linearly with the oxide thickness. As these simulations aim at studying the effect of the metal-oxide regrowth after performing an HF etch, the thickness of the substrate-oxide is kept constant at 0 nm.

Table S-VI: Simulated participation ratios and calculated microwave loss of various components (Si substrate, metal-substrate, metal-air, and substrate-air). Participation ratios are simulated for the reference sample. Participation ratios for these components are estimated as weighted average over capacitor (CR), meander (MR) and capacitor-meander (CMR) regions in the LE resonator. Asterix denotes the two possibilities, however, only one is likely present.

| | **CR** *F*(%) | **MR** *F*(%) | **CMR** *F*(%) | **Weighted** *F*(%) | **tanδ** **(x10⁻⁴)** | **Loss** **(x10⁻⁷)** |
|---|---|---|---|---|---|---|
| **Si*** | 92.0 | 90.1 | 91.1 | 91.6 | 0.0013 | 1.2* |
| **MS*** | $7.3 \times 10^{-3}$ | $166.5 \times 10^{-3}$ | $71.5 \times 10^{-3}$ | $35.4 \times 10^{-3}$ | 3.5 | 1.2* |
| **MA** | $0.7 \times 10^{-3}$ | $22.4 \times 10^{-3}$ | $9.1 \times 10^{-3}$ | $5.6 \times 10^{-3}$ | 99.0 | 5.3 |
| **SA** | $2.5 \times 10^{-3}$ | $60.1 \times 10^{-3}$ | $25.3 \times 10^{-3}$ | $11.9 \times 10^{-3}$ | 17.0 | 2.0 |
| | | | | Total loss: | | $8.5 \times 10^{-7}$ |
| | | | | **Q-factor:** | | $1.2 \times 10^{6}$ |

## APPENDIX E: SURFACE ROUGHNESS AFTER HF CLEAN

Surface roughness is measured on three blanket Nb samples (SR1-SR3) after various HF clean procedures. An increase of the surface roughness after the HF clean is clearly detected as shown in Table S-VII.

Table S-VII: Surface roughness study after the HF dip of blanket Nb films.

| Name | HF conc. (vol%) | Duration | RMS |
|---|---|---|---|
| Ref | N/A | N/A | 0.58 nm |
| SR1 | 0.9 | 90 s | < 1 nm |
| SR2 | 10 | 5 min | 6 nm |
| SR3 | 10 | 10 min | 9.4 nm |